\documentclass[11pt]{article}
\pdfoutput=1

\usepackage{hyperref}
\usepackage{comment}

\usepackage{graphicx}
\usepackage[utf8]{inputenc}
\usepackage{algorithm}
\usepackage{algpseudocode}
\usepackage{amsmath,amssymb,amsfonts}
\usepackage[dvipsnames]{xcolor}
\usepackage{mathtools}
\usepackage{subfig}

\usepackage{paralist}
\usepackage{threeparttable}
\usepackage{longtable} 
\usepackage{xspace}
\usepackage{tikz}
\usepackage[utf8]{inputenc}

\usetikzlibrary{shapes.misc}

\usepackage{calc} 
\usepackage{enumitem} 
\usepackage{comment}
\usepackage{longtable} 
\usepackage{multirow}
\usepackage{xspace}
\usepackage{tikz}
\usetikzlibrary{calc, arrows, decorations.markings}
\usetikzlibrary{shapes.misc}

\usepackage{import}
\usepackage{xifthen}
\usepackage{pdfpages}
\usepackage{transparent}
\usepackage{svg}

\graphicspath{{fig/}}
\tikzset{cross/.style={cross out, draw=black, minimum size=2*(#1-\pgflinewidth), inner sep=0pt, outer sep=0pt},
cross/.default={1pt}}
\newif\ifstatus
\statustrue 

\newcommand{\todolater}[1]{}

\DeclarePairedDelimiter\ceil{\lceil}{\rceil}

\usepackage{navigator}

\title{Towards Automated Homomorphic Encryption Parameter Selection with Fuzzy Logic and Linear Programming}

\author{Jos\'e Cabrero-Holgueras\thanks{jose.cabrero@alumnos.uc3m.es} \\ CERN, Universidad Carlos III de Madrid \and 
        Sergio Pastrana\thanks{spastran@inf.uc3m.es} \\ Universidad Carlos III de Madrid}

\date{October 2022}

\begin{document}

\maketitle





\begin{abstract}
Homomorphic Encryption (HE) is a set of powerful properties of certain cryptosystems that allow for privacy-preserving operation over the encrypted text. Still, HE is not widespread due to limitations in terms of efficiency and usability. Among the challenges of HE, scheme parametrization (i.e., the selection of appropriate parameters within the algorithms) is a relevant multi-faced problem. First, the parametrization needs to comply with a set of properties to guarantee the security of the underlying scheme. Second, parametrization requires a deep understanding of the low-level primitives since the parameters have a confronting impact on the precision, performance, and security of the scheme. Finally, the circuit to be executed influences, and it is influenced by, the parametrization. Thus, there is no general optimal selection of parameters, and this selection depends on the circuit and the scenario of the application. Currently, most of the existing HE frameworks require cryptographers to address these considerations manually. It requires a minimum of expertise acquired through a steep learning curve. In this paper, we propose a unified solution for the aforementioned challenges. Concretely, we present an expert system combining Fuzzy Logic and Linear Programming. The Fuzzy Logic Modules receive a user selection of high-level priorities for the security, efficiency, and performance of the cryptosystem. Based on these preferences, the expert system generates a Linear Programming Model that obtains optimal combinations of parameters by considering those priorities while preserving a minimum level of security for the cryptosystem. We conduct an extended evaluation where we show that an expert system generates optimal parameter selections that maintain user preferences without undergoing the inherent complexity of analyzing the circuit. 

\end{abstract}





\section{Introduction}
Modern society is highly dependent on technology. Tons of data are created, stored, and transmitted daily in cyberspace, which has become a critical infrastructure that demands proper security and privacy guarantees. The improvements in computation technologies (e.g., Deep Learning), networks, and storage have resulted in the advent of cloud-based services~\cite{cloud_privacy}. Indeed, most scenarios with Big Data processing and storage scenarios occur in a cloud-based infrastructure. Unfortunately, the nature of certain types of data often prevents its usage due to privacy concerns. It is the case with sensitive information such as medical~\cite{dl_medicine}, financial~\cite{dl_finance}, or cybersecurity~\cite{dl_cybersecurity}.

In that regard, the last decade has witnessed the growth of privacy-preserving computation techniques that enable processing in privacy-enforcing cloud-based scenarios~\cite{Gentry2009}. One such technique is Homomorphic Encryption (HE), which allows performing operations on encrypted ciphertexts preserving the changes on decryption~\cite{Regev2009}. For example, with HE, one could encrypt information, send it over to the cloud for processing, and receive the resulting ciphertext, guaranteeing that the exchange did not reveal any information~\cite{security_of_he}. The decrypted result contains the result of the computation as it was done over the non-encrypted information. Due to the powerful properties, HE is being adopted by modern technologies, e.g., to avoid privacy leakage of data shared on the Blockchain~\cite{qiao2022privacy}, for private computation within Deep Learning architectures~\cite{cabrero2021sok}, or for verifiable Association Rule Mining (ARM) in privacy-sensitive contexts~\cite{chen2022verifiable}. 
However, not only homomorphic encryption bears many challenges in terms of performance and scalability~\cite{lhe_slow_bootstrapping, fhew_slow_bootstrapping}, but also usability of the existing solutions~\cite{cabrero2021sok}. These challenges prevent its use when real-time data processing is required (e.g., in Machine Learning). Also, the complexity and expertise for elaborating efficient and secure code dissuade users from adopting them.

In recent years, a trend to overcome the challenges of HE has come in the form of compilers, frameworks, and APIs~\cite{Viand21}. Most frameworks and libraries provide a set of low-level primitives for computing with HE~\cite{microsoft_seal, ibm_helib}. However, these primitives are not aligned with the potential use programmers may give of HE, and they demand precise refinement for real applications. 
On top of these libraries, some frameworks provide an abstraction layer to HE functions, without the need to deal with the inner workings of the algorithms. 
For example, compilers such as CHET~\cite{chet, pyeva} transform a given circuit (defined as a simpler cleartext alternative) into its HE counterpart for use on encrypted data. Despite being a promising direction to increase HE adoption at scale, these frameworks and compilers are still in their early days and sometimes require the user to have expertise and knowledge to properly use them~\cite{cabrero2021sok}.

As we outline in Section~\ref{sec:related_work}, one of the main challenges of HE schemes, which is not addressed in existing frameworks, is the selection of a set of critical parameters, i.e. the parametrization. In a nutshell, the parameter selection is complex since there are circular relations between parameters, and with the underlying circuit. Also, the combination of different values for these parameters results in a trade-off between three essential features of privacy-preserving scenarios, i.e., security (having more robust encryption), performance (requiring a reasonable amount of computational resources), and precision (limiting to an affordable amount of noise in the final result). Furthermore, in order to understand the consequences, the user needs detailed knowledge about the low-level primitives. We provide a gentle introduction to HE and further break down these problems in Section~\ref{sec:background-he}. 
The HE Standard provides minimum settings to establish secure parameters in Homomorphic Encryption~\cite{he_standard}. Similar to classical cryptography, where the selection of critical parameters (such as RSA key size and AES block size) is not left to end-users or programmers, for a widespread adoption of HE, solutions shall not leave these decisions to users.  Still, it is needed to provide degrees of freedom for the user to determine parameters such as security, performance, or precision, since these might depend on the scenario of application of the circuit.

In this paper, we propose an expert system that allows the selection of an optimal set of parameters for HE. Given an input circuit, our system leverages Linear Programming to select a close-to-optimal solution given a set of intrinsic rules to the HE scheme. The characteristics and constraints of HE parameter selection enable their representation as a Linear Programming Model. Moreover, due to the potential parameter choices and their confronting impact on security, efficiency, and performance, the system allows users to guide the selection around these constraints. Concretely, the system uses Fuzzy Logic to select the best options based on the criteria defined by the user. We believe that this system would be of benefit to programmers not familiar with the inner cryptography of HE schemes.
Concretely, we make the following contributions:
\begin{enumerate}
    \item We provide a theoretical introduction to the concepts of Levelled Homomorphic Encryption (LHE), which is a widely use approach since it is practical in terms of accuracy and performance. We motivate the problem of parametrization, and explain what are the main challenges faced by programmers when dealing with LHE in Section~\ref{sec:background-he}.
    \item We propose and detail an expert-system that combines Fuzzy Logic and Linear Programming to automate parameter selection (Section~\ref{sec:model}). The automation considers the different implications of parameter selection, and guarantees compliance with the HE Standard~\cite{he_standard}.
    \item We conduct an evaluation for the performance, precision and security of the system under different scenarios of applications, and executing circuits that entail different workloads on a benchmark set using the well-known HE library Lattigo~\cite{lattigo} (Section~\ref{sec:evaluation}).
\end{enumerate}

We believe that the proposed system considerably contributes to the literature by providing a new mechanism to solve a recurrent problem within the Homomorphic Encryption, i.e., the optimal selection of parameters, by reducing the expertise required for this task. To allow for reproducibility and foster its application into existing frameworks, we further detail the system and provide an open-source prototype in our public repository.\footnote{\url{https://github.com/jcabrero/flp_he_parametrization}}

\section{Background}
This section provides a theoretical background of the different concepts tackled in this paper to enable a proper understanding of the complete expert system presented.

\subsection{Linear Programming}


Linear Programming (LP) is an optimization technique that obtains -if existent- the optimal solution to a model, which is defined by an objective function, the decision variables, and a set of constraints~\cite{linprog1}. Concretely, LP attempts to maximize or minimize the objective function. The objective function shows the overall benefit or cost of decision variables. In the objective function, $x$ represents the decision variables and $c$ are the coefficients as follows: $$ max~z = c^Tx~|~min~z = c^Tx$$
The objective function is ruled by the constraints that are usually represented as a matrix $A$ and a vector of resources $b$ as follows: $$ Ax \leq b$$$$x \geq 0$$

In LP tasks, the constraints define a region of all feasible solutions (i.e., the solutions to the objective function that fulfill the constraints). Optimal solutions usually lie in the extreme points of the feasible region. The Looseness FL Module, described in the following section, generates coefficients that modify the constraints of the feasible region so that extreme points vary.

LP tasks are solved with methodologies widely falling under two categories, i.e., Simplex-based methods and Interior Point Methods. Simplex-based methods work on the polyhedron constructed with the decision variables and bounded by the constraints~\cite{linprog_simplex}. In Interior Point Methods, the algorithms work with the feasible region and travel the extreme points to find the optimal solutions~\cite{linprog_interior_point}. Although a priori the methods have different computational complexities, their practical efficiencies are similar nowadays.

\subsection{Fuzzy Logic}
Fuzzy Logic (FL) is a control system technique aimed at systems working with a degree of uncertainty. In particular, FL profits from language rules and uses them to express control system values. The benefit of FL is the ability to represent a value within a threshold. In FL, there exist two predominant solvers, Mandiani and Sugeno. For this paper, we use Mandiani as it delivers more detailed explanations of the decision variable relations.

During the fuzzy inference process, a set of crisp antecedents (i.e., crisp values that we know as truth) are transformed into fuzzy sets to relate them logically and obtain a crisp consequent. The complete process is the following: first, the antecedents are \textit{fuzzified}. The \textit{fuzzification} analyzes the degree of membership of the crisp value to membership functions. In the second phase, the fuzzy inference process relates the different fuzzy values (i.e., belonging to the membership functions) through boolean logic statements. The result of the fuzzy inference process is a fuzzy consequent, which is \textit{deffuzified} with some predefined metric (i.e., usually the area of the centroid) to obtain the final value.  

\subsection{Homomorphic Encryption}\label{sec:background-he}
Homomorphic Encryption (HE) is a property of an encryption scheme that allows it to operate over the encrypted operands (ciphertexts) while preserving the actual operation results on the underlying cleartext operands. Homomorphic Encryption regained popularity when Gentry~\cite{Gentry2009} proved Fully Homomorphic Encryption (FHE) feasible through bootstrapping (i.e., unbounded computation). Despite theoretical advances, in practical terms, bootstrapping remains relatively inefficient. Thus, Levelled Homomorphic Encryption (LHE) schemes, which only allow a limited number of multiplications, are more widespread due to their efficiency. In this paper, we focus on LHE schemes. 

Most modern homomorphic encryption schemes rely on the Learning With Errors (LWE)~\cite{Regev2009} problem. The cryptographic strength of the LWE problem comes from the solution of a system of linear equations with an introduced error. LWE is lattice-based and it is considered quantum-resilient. In particular, we focus on one of the most popular schemes, i.e., CKKS~\cite{ckks}. This scheme introduces floating point operation capabilities on the ciphertexts. Furthermore, it offers Packed Homomorphic Encryption, which permits encoding a vector of plaintext values within a single ciphertext allowing for Single Operation Multiple Data (SIMD) operations over all the entries. SIMD operation results in a  more efficient computation but introduces further complexity for algorithm design and parameter selection~\cite{Cabrero2022Towards}.

\subsubsection{HE using Learning With Errors schemes}

An LWE-based HE scheme allows encrypting a message $m$ using a secret $s$ both represented as polynomials in $Z^N_Q$, i.e., the set of polynomials of degree $N$ with modulus $Q$ (with $N, Q \in \mathbb{Z}^+$). It also requires setting a parameter $\sigma$ for the standard deviation of a random distribution $\chi(\mu, \sigma)$, used to sample random polynomials $a \in Z^N_Q$. As we detail later, selecting appropriate values for the parameter set ($N$,$Q$, and $\sigma$) is complex, and indeed is the problem that we address in this paper.

The scheme hides a secret vector $s \in \mathbb{Z}_Q^N$ by sampling a random polynomial whose solution is $s$ and adding random noise $e \in Z$ sampled from a random distribution $\chi(\mu, \sigma)$. From that, it outputs a tuple of polynomials $(a, b)$, with:

$$a = c_0 + c_1x + c_2x^2 + ... + c_Nx^N \mod Q.$$ 
$$b=\langle a, s \rangle + m + e$$


In short, LWE randomly samples a polynomial $a$ of degree $N$ modulo $Q$. Then, thanks to $s$, it hides $m$ in the polynomial by adding some error $e$. The result of this hiding is $b$, which with $a$, both together conform to the ciphertext.

The robustness of LWE schemes relies on two problems that try to obtain the secret vector $s$ from the tuple $(a, b)$. The Search-LWE problem tries to find the solution to the polynomial to recover $s$, and the Decision-LWE problem tries to distinguish a polynomial hiding a message from a random sample. 

In LHE schemes, two special operations, relinearization and modulo switching follow every multiplication operation.
The multiplication of two polynomials of degree $N$ increases the degree (e.g., $x^N * x^N  = x^{2N}$). The relinearization uses the encrypted square of the secret $s$ to reduce the degree. 
The modulo-switching operation aims to reduce the noise introduced with each multiplication. For this process, the polynomial modulus $Q$ is defined according to the Chinese Remainder Theorem as the multiplication of $\mathcal{O}_{\mathcal{D}}$ smaller moduli $q_i$, i.e., $Q = \prod_{i = 0}^{\mathcal{O}_{\mathcal{D}}} q_i$. The modulo switching sometimes has a particular behavior in certain libraries. If the scale has not grown significantly, there is no need to rescale and perform modulo switching. In the same way, if the scale has grown significantly, the rescaling may be performed twice. In our experience, these cases only occur with small modulus and are very rare, so in this paper, we assume only the case where each rescaling uses one modulo and is always performed. 

The term $\mathcal{O}_{\mathcal{D}}$, which corresponds with the multiplication depth of the circuit, defines the minimum number of moduli $q_i$ needed to operate. 
For simplicity (and consistency with HE libraries), in the remainder we use $logQ$, $logN$ and $logq_i$ to refer to the bit counts of these numbers (e.g., $logQ = \ceil{log_2(Q)}$). 

In contrast to general LWE schemes, LWE-based HE schemes require a careful parameter selection, as we cover next. 

\subsubsection{LWE Parametrization for LHE}~\label{subsec:lwe_parametrization}

\sloppy In general, HE schemes aim at reducing the noise introduced by operations, so circuits allow more operations~\cite{security_of_he, security_of_he2}. In this regard, the Homomorphic Encryption Security Standard~\cite{he_standard} provides a set of guidelines for the secure selection of these parameters. In this paper, we adhere to those guidelines and adopt them strictly in the design of the proposed system.

The process of assigning a parametrization $(N, Q, \sigma)$ to a circuit $\mathcal{C}$ for its execution under LHE poses many challenges.
In most cases, there is a unidirectional dependence of the parametrization $(N, Q, \sigma)$ on the circuit $\mathcal{C}$. 

In such case, the first step is to extract the circuit constraints, precision $p$, multiplication depth $\mathcal{O}_{\mathcal{D}}$ and maximum vector length $|v|_{max}$. 
The precision $p$ is the number of bits needed to represent the real part of the largest floating point number in the computation $p =\ceil{log_2(d_{max})}$. Note that it does not only affect input variables but also intermediate and resulting values. 
The multiplication depth $\mathcal{O}_{\mathcal{D}}$ is the maximum number of consecutive multiplications that occur over a ciphertext in the circuit $\mathcal{C}$. 
The maximum vector length $|v|_{max}$ is the length of the largest vector that we aim to encode in a single ciphertext. Most HE libraries limit the slots to $2^{15}$ for CKKS, and thus it may be necessary to represent a vector in multiple ciphertexts. This step may introduce additional computation rounds and multiplication depth~\cite{Cabrero2022Towards}.

After obtaining these values ($p$,$\mathcal{O}_{\mathcal{D}}$ and $|v|_{max}$), the second step is to select the parameters $(N, Q, \sigma)$. 

To reduce the amount of noise, in most cases, and according to the HE Standard, $\sigma \approx 3.2$.  Increasing $\sigma$ improves the security but reduces the precision during decryption. That is due to the introduction of more noise per operation. 

Next, the chain of moduli $q_i$ is selected, which includes at least $\mathcal{O}_{\mathcal{D}}$ $logq_i$-bit prime numbers, and whose product defines the polynomial module $Q$. 
Finally, according to the guidelines of the Homomorphic Encryption Security Standard~\cite{he_standard}, each $N$ allows for a maximum budget or amount of moduli bits, represented as $\mathcal{B}_{N, \lambda, \mathcal{T}}$. More formally, for each value of $N$, for a security type $\mathcal{T} \in \{classical, quantum\}$ and for a security parameter $\lambda \in \{128, 192, 256\}$ the standard defines a maximum budget that must be lower than the sum of the selected $logq_i$ ($\mathcal{B}_{N, \lambda, \mathcal{T}} >= \sum_{i=0}^{\mathcal{O}_{\mathcal{D}}} logq_i$). Therefore, for a circuit parametrization to be secure, one would need to compute the sum of all $logq_i$ and then compare it to the budget $\mathcal{B}_{N, \lambda, \mathcal{T}}$ established to obtain a $N$. 

At the same time, $N$ shall compare with the maximum vector length $|v|_{max}$. If $\dfrac{N}{2} >= |v|_{max}$, then the largest vector fits within the ciphertext slots of the chosen parameter set. 
Furthermore, to ensure that the maximum precision does not overflow, the distance between $q_0$ (known as the \textit{special prime}) and the rest of the moduli $q_i$ has to be $p$ following $logq_0 = logq_i + p~\forall~i > 0$. We note that in practice only $logq_1$ needs to preserve this property but to maintain the cost of rotations negligible, all $logq_i$ must share the same bit count.

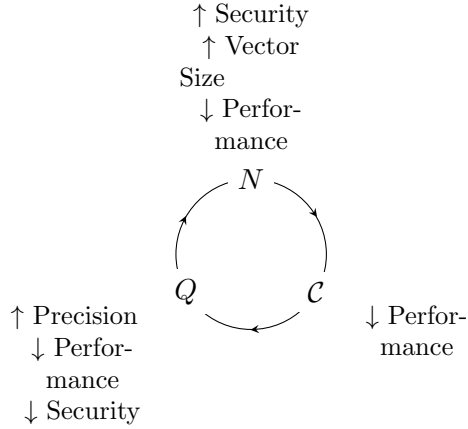
\begin{figure}[htb]
  \centering
    \begin{tikzpicture}[decoration = {markings,
        mark = between positions 0.1 and 1 step 0.3333 with {\arrow[>=stealth]{<}}
      },
      every label/.style={text width=0.75in,align=center,font={\small}}
      ]
      \draw[postaction = decorate] (0, 0) circle [radius = 1cm];
      \path (90 :1cm) node[fill=white,label=90:{$\uparrow$~Security $\uparrow$~Vector Size\newline $\downarrow$~Performance}]{$N$}
            (330:1cm) node[fill=white,label=-10:{$\downarrow$~Performance}]{$\mathcal{C}$}
            (210:1cm) node[fill=white,label=190:{$\uparrow$~Precision\newline $\downarrow$~Performance $\downarrow$~Security}]{$Q$};
    
    \end{tikzpicture}
  \caption{Circular dependencies in HE parametrization between polynomial degree $N$, polynomial modulus $Q$ and circuit programming $C$. The diagram also depicts the implications of each parameter in the growing ($\uparrow$) or decreasing ($\downarrow$) of each variable. }
  \label{fig:circular_rel}
\end{figure}

However, the above reasoning misses two important factors in the selection of parameters.
First, if $|v|_{max}$ does not fit within the biggest vector, the plaintext vector is represented by multiple smaller ciphertexts (i.e., involving additional multiplication depth for each rotation, thus modifying the circuit). 
Also, some algorithms rely on the parameter $N$ for their execution. For example, the vector aggregation algorithm (i.e., sums all the entries of a vector) performs $logN$ consecutive multiplications. Then, this behavior involves a different multiplication depth $\mathcal{O}_\mathcal{D}$ per each possible $N$ value (in this paper, we denote this relation between the multiplication depth and $N$ as $\mathcal{O}_\mathcal{D}^N$). Using such algorithms requires an iterative parametrization process since the different choices of $(N, Q, \sigma)$ may introduce changes in the circuit $C$, which in turn might require re-definitions of the parameters as represented by Figure~\ref{fig:circular_rel}. One potential solution is to keep fixed one or two of the parameters. However, this can bias the parametrization and make it less efficient. 

The second factor not mentioned before is related to the impact of the parameter selection on the output. Indeed, each parameter selection involves a tradeoff between precision, security, and performance. For example, increasing  $N$ increases the security since a larger polynomial hides the secret, but it becomes less efficient since it requires operations over a higher degree polynomial. Similarly, higher values of $Q$ allow for larger multiplication depth and better precision. However, the smaller the value $Q$, the more secure and efficient the encryption scheme is.  

\vspace{0.3cm}


\section{Related Work} ~\label{sec:related_work}

In recent years, some proposals have looked at the affordability of HE techniques for end-users, presenting HE protocols as application programming interfaces (API) or compilers~\cite{Viand21}. Compilers and APIs are software pieces that aim to provide a simple functionality for end programmers while hiding the complexity employed in the logic of programs. In HE, compilers and APIs allow the users to access abstracted representations of low-level instructions with reduced complexity and within standard programming languages. The software generates a logical coupling between the high-level model and the underlying HE circuit, concealing the particularities of the technique. However, most compilers still do not provide an automated parameter selection, requiring the user to select them manually. nGraph HE~\cite{ngraph, ngraph2, mp2ml} by Intel delivers an API to HE focused on its use for Deep Learning. However, the user must manually perform the parameter selection. Also, with the complexity hidden from the user, understanding factors such as the depth of the circuit remains intricate.
PySyft~\cite{pysyft} is an open-source initiative to bring privacy-preserving techniques closer to end-users by providing a friendly Python interface. As part of their implementation, they have ported Microsoft SEAL~\cite{microsoft_seal} to work on Python on TenSEAL. Similar to nGraph, the parameter selectionin PySyft is not automated and has to be done by the user. 
The more advanced solutions are those of CHET~\cite{chet}, EVA~\cite{pyeva} and Concrete~\cite{concrete}.  
The compiler CHET~\cite{chet}, and the API EVA~\cite{pyeva} are both developed by Microsoft as an abstraction for the HE library SEAL~\cite{microsoft_seal}. 
While they provide a relatively automated parameter selection, the behavior is too simplistic and thus, not optimal. It chooses a default scale and using it as many times as needed by the multiplication depth, and then setting the rest of the parameters based on the total $logQ$. As a difference to our proposal, EVA requires the user to input the maximum precision for the real scale. In our case, we decide it through a user-defined value for precision priority. 
Concrete is a HE library created by ZAMA as an evolution of the TFHE~\cite{tfhe} scheme. Concrete has developed different levels of abstraction for end-users to adapt to their knowledge. In the lower layers, the user needs to provide the parameters, but in the upper layers, the programming interface takes care of that burden. Unfortunately, the comparison is not direct since Concrete relies on TFHE (i.e., boolean arithmetic),  and our proposal leans on CKKS (i.e., floating point arithmetic).

\vspace{0.5cm}

\noindent\textbf{Take-away.} Overall, a proper setup of FHE schemes such as CKKS~\cite{ckks} highly depends on the optimal selection of parameters. Unfortunately, such optimal selection is a complex task since parameters depend on each other and the underlying circuit. Moreover, the selection has serious impacts on the security, efficiency, and precision of the output. Balancing the trade-off of these features depends on the scenario of the application. 
Existing FHE frameworks usually leave the selection of parameters to the user, which is time-consuming and requires expertise in cryptography. Thus, in this paper, we propose an expert system that leverages Linear Programming and Fuzzy Logic to assist in parameter selection and overcomes these challenges.

\section{System Model} 
\label{sec:model}

\begin{figure}[!htbp]
    \centering
    \includegraphics[width=\columnwidth]{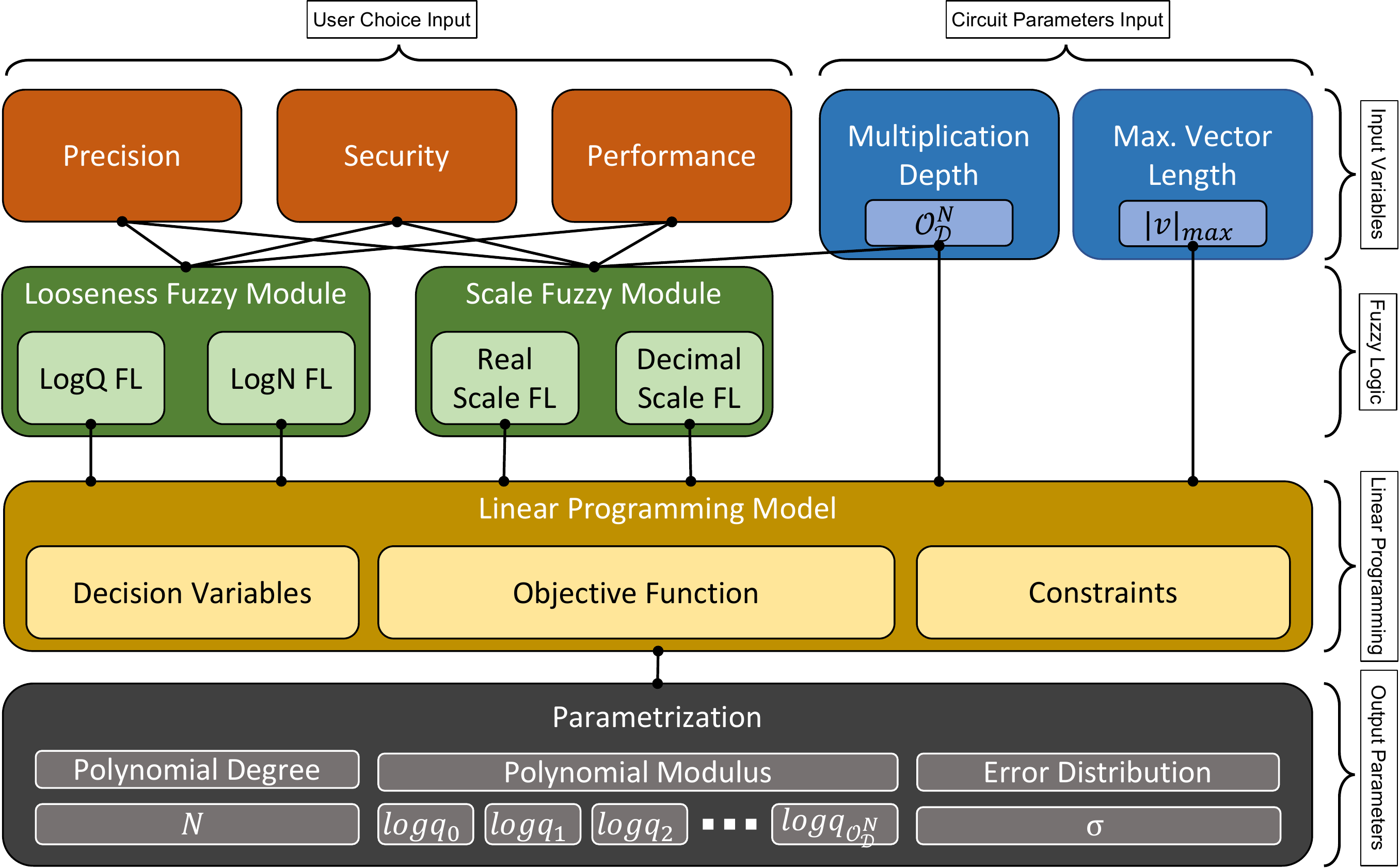}
    \caption{General methodology of the proposed system}
    \label{fig:architecture}%
\end{figure}

In the previous section we described the process for parameter setup in HE. We also discussed the difficulty of manually establishing appropriate values in absence of expertise on the inner cryptographic working of the scheme. Also, the complexity of the LWE scheme makes that, even a human with expertise might fail to provide an optimal set of parameters, due to circular dependencies among them, and with the circuit being operated. Also, as discussed before, the selection of parameters has an important impact on the performance, security and precision of the scheme. These are confronting goals, and its setup depends on the scenario of application (e.g., one would prefer to have a more secure scheme, even at the cost of losing precision or performance, and viceversa). 
Accordingly, we propose a system that automatically finds the optimal parameters and, to make the system more flexible, 
asks the user to choose a priority level for these confronting features. 

The methodology is depicted in Figure~\ref{fig:architecture}. It receives as input the level of priority for each of the aforementioned features, i.e., security, precision and performance, in a score from 0 to 10. We note, however, that stating priority of security of 0 means putting \textit{the lowest priority} on secure parameters (i.e., reducing $N$ and increasing $Q$). But in no way the system will generate an insecure set of parameters, as established by the HE Standard~\cite{he_standard}. 
The system applies Fuzzy Logic, which permits to make fuzzy decisions on parameters while establishing fuzzy rules on the input values. The output of the Fuzzy Logic, together with information regarding the circuit (i.e., the multiplication depth, and the maximum vector length) are used to define the Linear Programming (LP) model. Concretely, it defines the LP constraints, decision variables and objective function. The LP model allows to balance the trade-off of each of the value choices for the selected parameters.
The remainder of this section describes further the Fuzzy Logic and the Linear Programming processes.



\subsection{Fuzzy Logic Initialization}
In this section, we elaborate on the two different Fuzzy Logic Modules used to generate the coefficients used to design the Linear Programming model. Concretely, the system relies on two fuzzy modules, i.e., the \textit{Scale FL} and the \textit{Looseness FL}. The Scale Fuzzy Module generates the maximum and minimum values of the polynomial modulus $logQ$ fulfilling the user input requirements in terms of security, performance, and precision. Then, to make the LP task more flexible, the Looseness Fuzzy Module expands the interval of valid values for the LP task to adjust the final parameters in an unconstrained interval. The combination of these two models outputs the coefficients used to define the actual LP model values that later define the constraints and objective function used in linear programming. We next explain these two models. 

\subsubsection{Scale Fuzzy Module}
The Scale Fuzzy Module produces two coefficients $k_{real}$ and $k_{dec}$ using two independent processes, i.e., the Real Scale Fuzzy Logic and the Decimal Scale Fuzzy Logic. The LP tasks use these coefficients to determine the bit length of each polynomial moduli $logq_i$. On the one hand, the Decimal Scale represents the total bit length value of $logq_i$. On the other hand, the Real Scale represents the value of precision $p$ needed (i.e., the number of bits to represent the real part of a decimal number).

\begin{figure}[!htbp]
    \centering
    \includegraphics[width=.6\columnwidth]{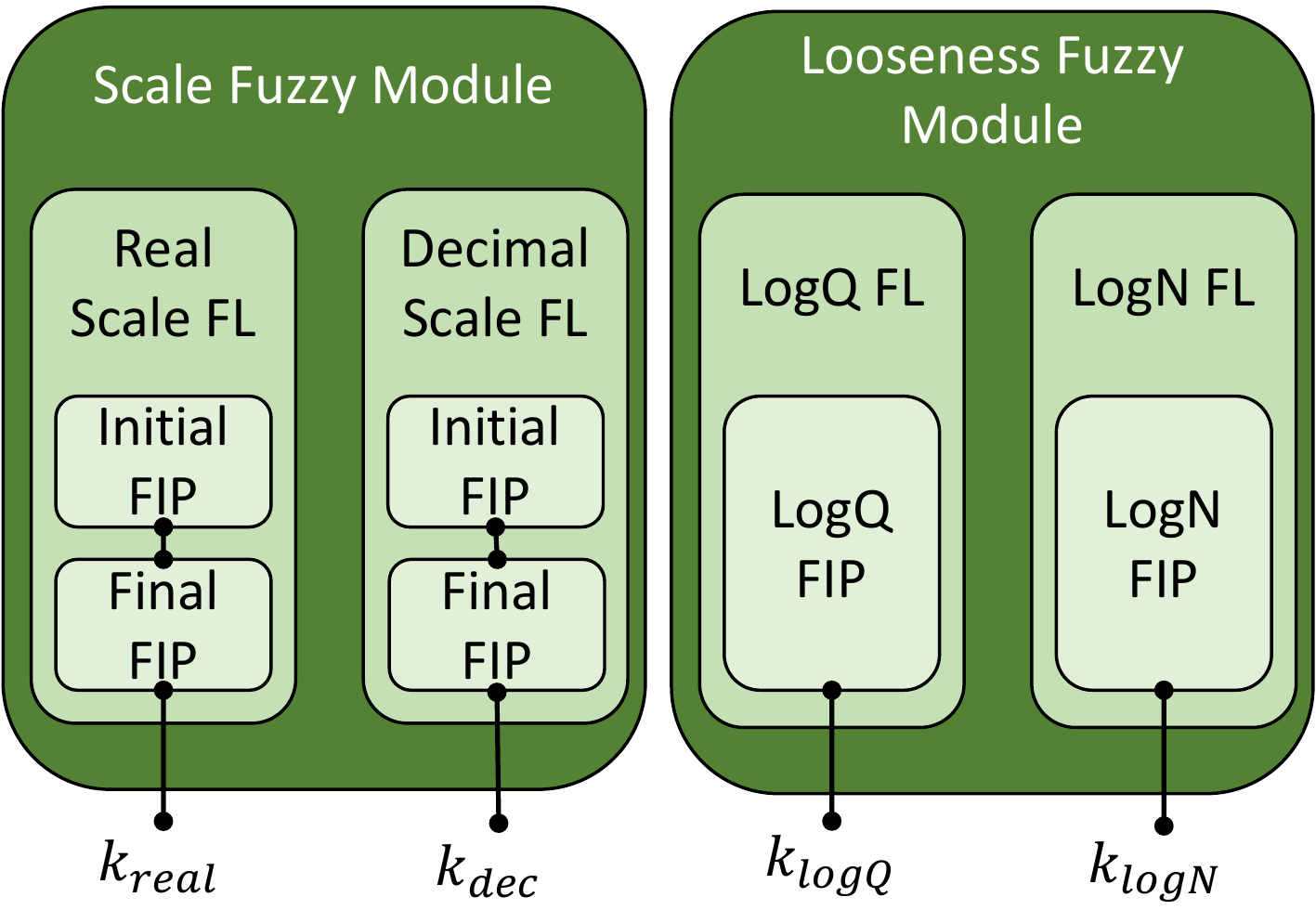}
    \caption{Overall hierarchization of the different Fuzzy Inference Processes (FIPs) carried in the Fuzzy Logic phase. For the Scale Fuzzy Module, there are two Fuzzy Logics, each of which has two separate FIP (i.e., Initial FIP and Final FIP. In the case of the Looseness Fuzzy Module, there is two Fuzzy Logic with a single FIP.}
    \label{fig:scale_fl_diag}%
\end{figure}

Both the Real and Decimal Scale FL consist of two consecutive Fuzzy Inference Processes (FIP), Initial FIP and Final FIP. The initial FIP only consider the priorities of the users for security, performance and accuracy. The final FIP refines the previous output by also considering information about the circuit (i.e., the multiplication depth). Figure~\ref{fig:scale_fl_diag} depicts how the chaining of the different FIPs within each Fuzzy Logic occurs, which are detailed next.

The Real and Decimal Scale FL share four antecedents and one consequent. The first three antecedents are the inputs from the user, i.e., the valuation (from 0 to 10) of the importance given to precision, performance, and security. The fourth antecedent is the \textit{multiplication depth} $\mathcal{O}_\mathcal{D}^N$, that shall be obtained from the circuit. Each FIP produces one consequent (coefficient), i.e., the real and decimal scale ($k_{real}$ and $k_{dec}$ respectively).

\begin{figure*}[!htbp]
    \centering
    \subfloat[\centering Real Scale]{{\includegraphics[width=0.5\textwidth]{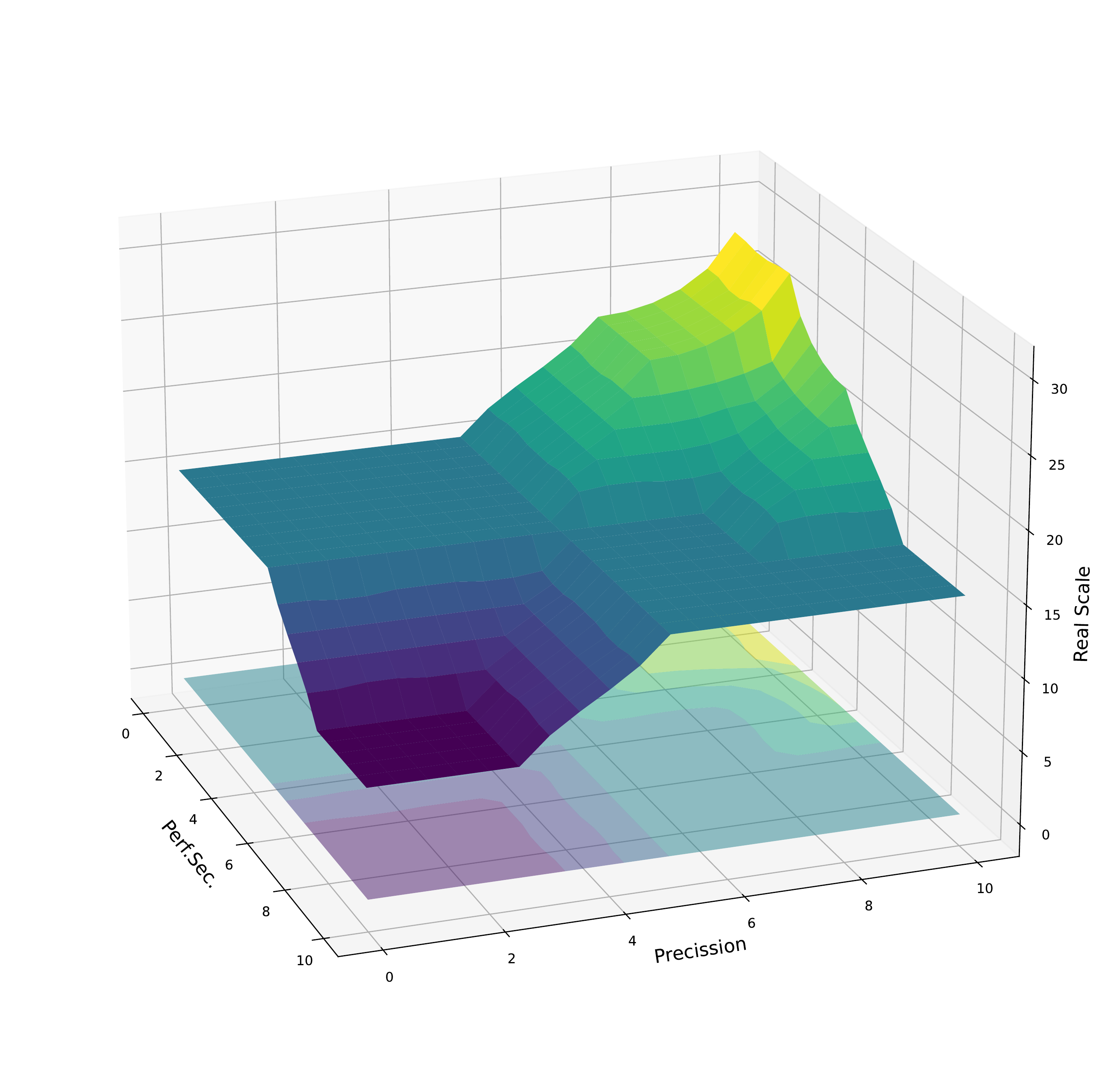} }}%
    \subfloat[\centering Decimal Scale]{{\includegraphics[width=0.5\textwidth]{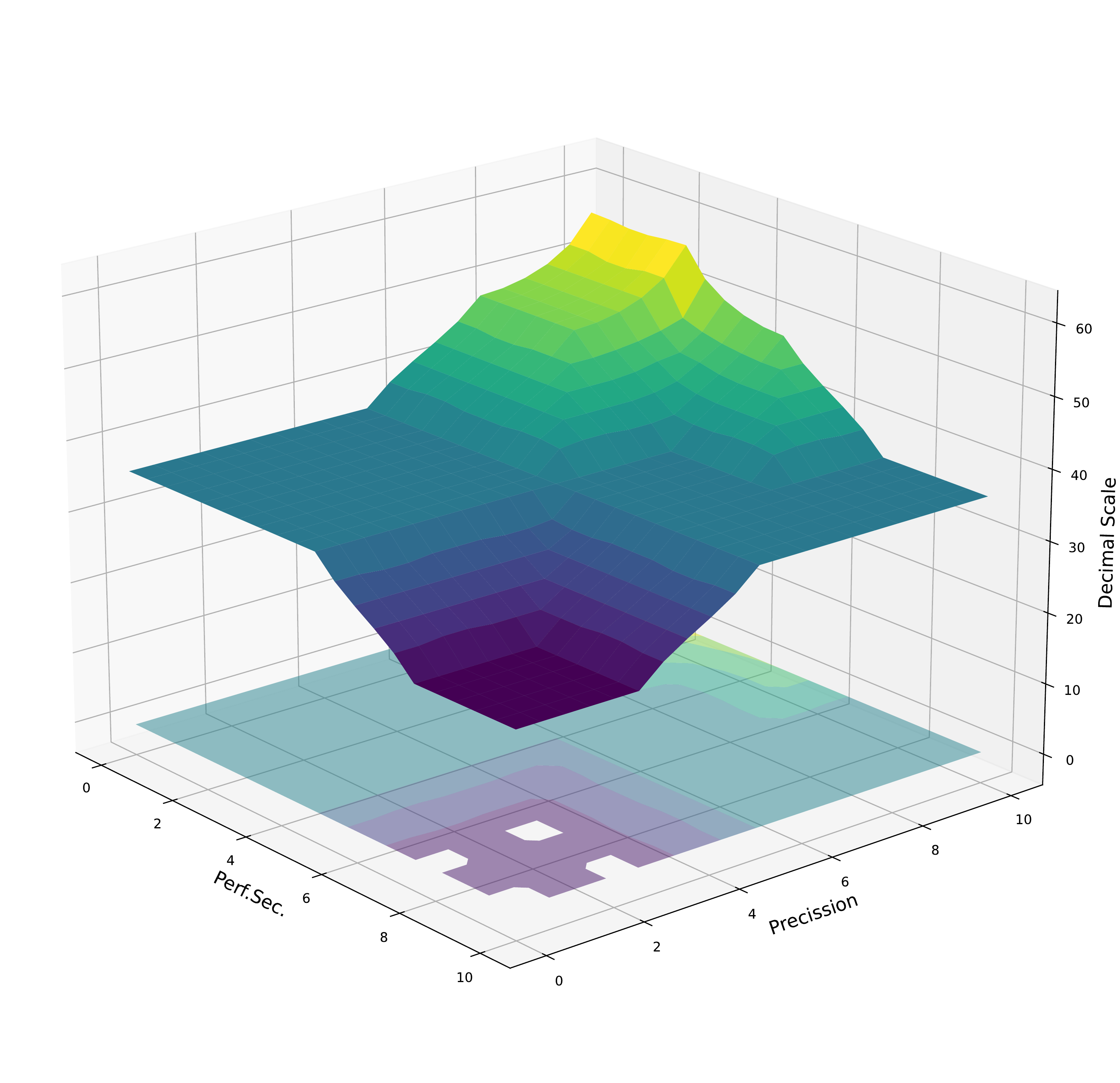} }}%
    \caption{Initial Fuzzy Inference Processes (FIP) for Real and Decimal Scale estimation obtention based on \textit{precision}, \textit{performance} and \textit{security} as antecedents. Given the aligning goals of \textit{performance} and \textit{security}, these are combined within the same metric. }
    \label{fig:scale_fl_initial}%
\end{figure*}

As mentioned before, for each of the FL processes, two consecutive FIPs are applied. These use five membership functions to describe the antecedents and consequent as \textit{very low}, \textit{low}, \textit{medium}, \textit{high} and \textit{very high}. For more details on the intervals for each membership functions, we refer the reader to Table~\ref{tab:intervals}. Splitting the Fuzzy Logic into two separate FIP significantly reduces the number of rules $5^2 << 5^3$. Also, it produces an easier understanding and relations of antecedents based on the logic rules.

The first FIP for both Real and Decimal Scale FL uses as antecedents the \textit{precision}, \textit{performance} and \textit{security} to compute an initial estimation of the Real and Decimal Scale as consequents. We note that, for the computation of $logq_i$, \textit{performance} and \textit{security} are aligning goals. Thus we compute the maximum of these two to compete with the \textit{precision}\footnote{We use the max value since our preliminary experimentation showed that using other metrics such as the average lead to less consistent results with the user inputs.}. Figure~\ref{fig:scale_fl_initial} shows the overall Real and Decimal Scale estimations according to the user inputs. The main variations among the Real and Decimal Scales are in the resulting ranges (i.e., the real scale follows a shorter range than the decimal scale). The graphs show how, for \textit{medium} values of precision, performance, and security (i.e., between 4 and 6), the FIP averages the result obtaining medium values of scale (i.e., around 15 for the scale and 30 for the decimal coefficient). However, in extreme contrary cases (e.g., high precision and low performance/security), the resulting decimal and scale values are outermost. We note that, in our implementation, we favor precision over performance and security as it may yield results useless otherwise. Therefore, we see a lower decrease in precision suffered by the surface but a larger decrease in performance and precision. 

\begin{figure*}[!htbp]
    \centering
    \subfloat[\centering Real Scale]{{\includegraphics[width=0.5\columnwidth]{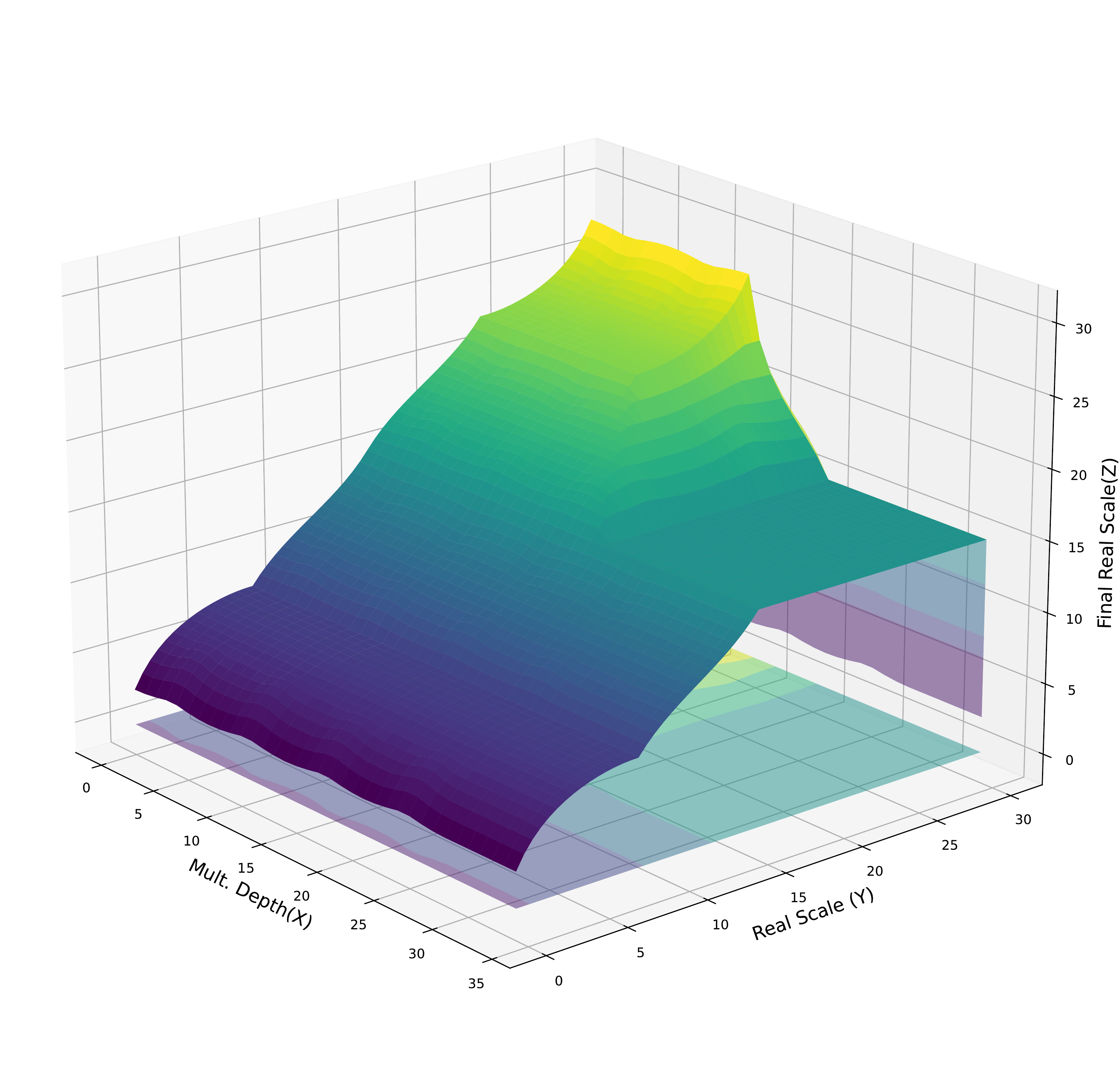} }}%
    \subfloat[\centering Decimal Scale]{{\includegraphics[width=0.5\columnwidth]{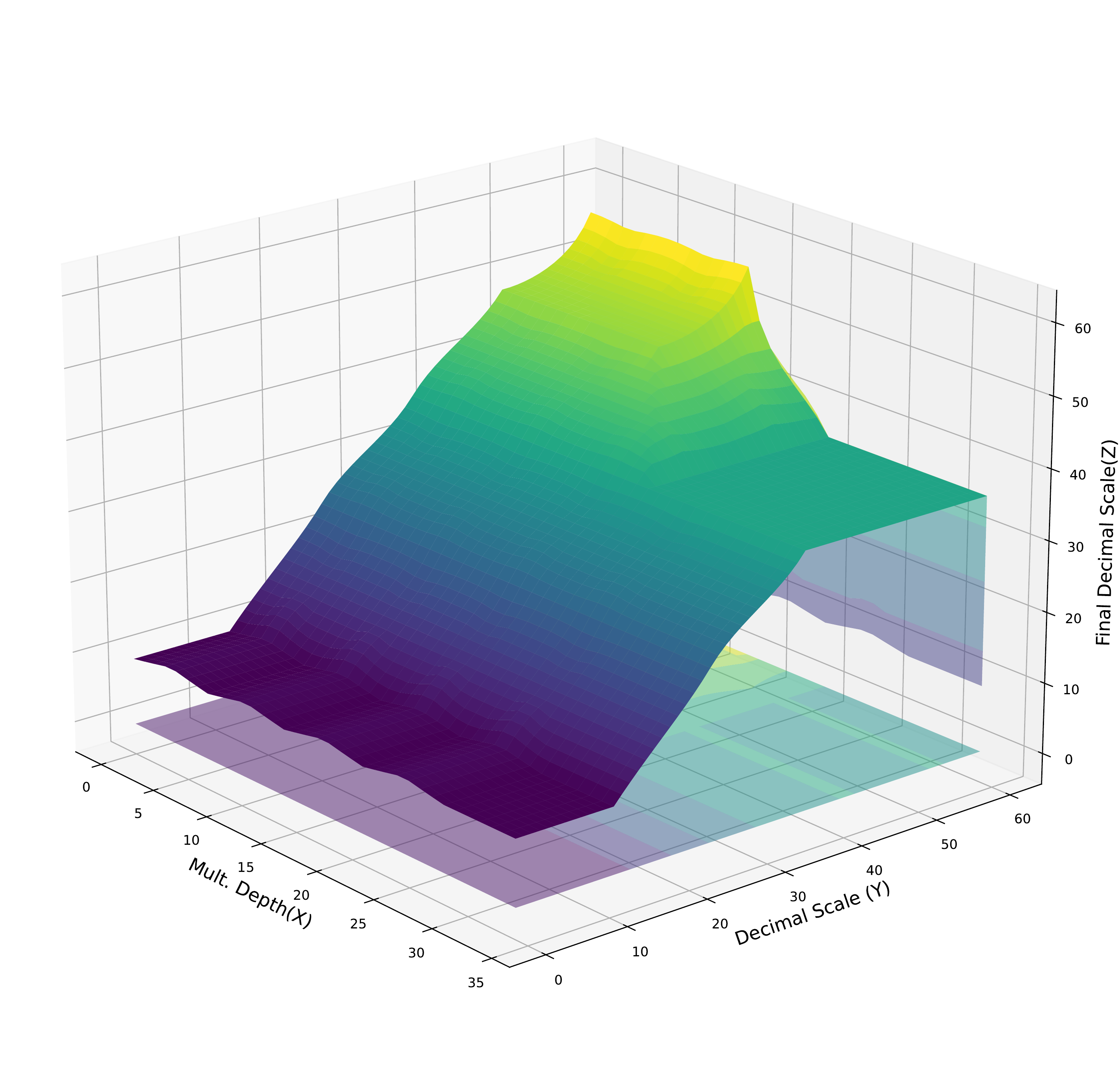} }}%
    \caption{Final Fuzzy Inference Processes (FIP) for Real and Decimal Scale coefficients obtention based on \textit{initial Real or Decimal Scale Estimation} and \textit{Multiplication Depth}. }    \label{fig:scale_fl_final}%
\end{figure*}

Despite the user's choices, the circuit restrains the permitted values for the real and decimal scales. For example, if the user chooses precision as its goal, the Real and Decimal Scales ($k_{real}$ and $k_{dec}$) will grow significantly. If the circuit also has a high multiplication depth, the maximum budget allowed within $N=15~\&~\lambda = 128$ may not fit enough bits of $logq_i$ and thus result in a non-solvable LP task. 
Accordingly,  the second FIP rationalizes the real and decimal scale coefficients by either maintaining or reducing them, based on the \textit{multiplication depth} of the circuit. Figure~\ref{fig:scale_fl_final} shows the surface plots of the results for the real and decimal scales. We can see how for low values of Multiplication Depth (i.e., simpler circuits), the coefficients are maintained from the previous FIP. However, for larger values of the estimated scales and multiplication depth, we observe a plateau where the final values are rationalized and reduced. In this way, the resulting coefficients fit within the maximum security budget. The main difference between the Real and Decimal Scale processes is for \textit{low} predicted scales. Wherein the Real Scale tends to increase its final value even for predicted low values (so it does not decrement precision), the decimal scale is kept constant across the \textit{very-low} values (i.e., from 0 to 20).

\subsubsection{Looseness Fuzzy Logic}
The Looseness FL provides margins that capture the need for flexibility of the LP task, and while preserving user priorities, it permits a range where the parametrizations still remain optimal. The intervals provide flexible room for the LP task to extract the more optimal values in terms of performance (i.e., choosing a more optimal $logN$) and precision (i.e., profiting from the $logQ$ budget). It benefits from the properties of the centroid defuzzification (i.e., it never takes extreme values) to establish flexible intervals for the LP task to work. 

This module extracts two different coefficients, $k_{logN}$ and $k_{logQ}$, that scale the polynomial degree $N$ and polynomial modulus $Q$, respectively. Each of them uses a different FIP. The FIP for $k_{logN}$ uses two antecedents, i.e., \textit{performance} and \textit{security}, since increasing $N$ improves security but reduces performance. The FIP for $k_{logQ}$ uses three antecedents, i.e., \textit{precision}, \textit{performance} and \textit{security}. However, similar to the Scale FL, as \textit{performance} and \textit{security} are aligning goals, we take the maximum one. 

\begin{figure}[!htbp]
    \centering
    \includegraphics[width=\columnwidth]{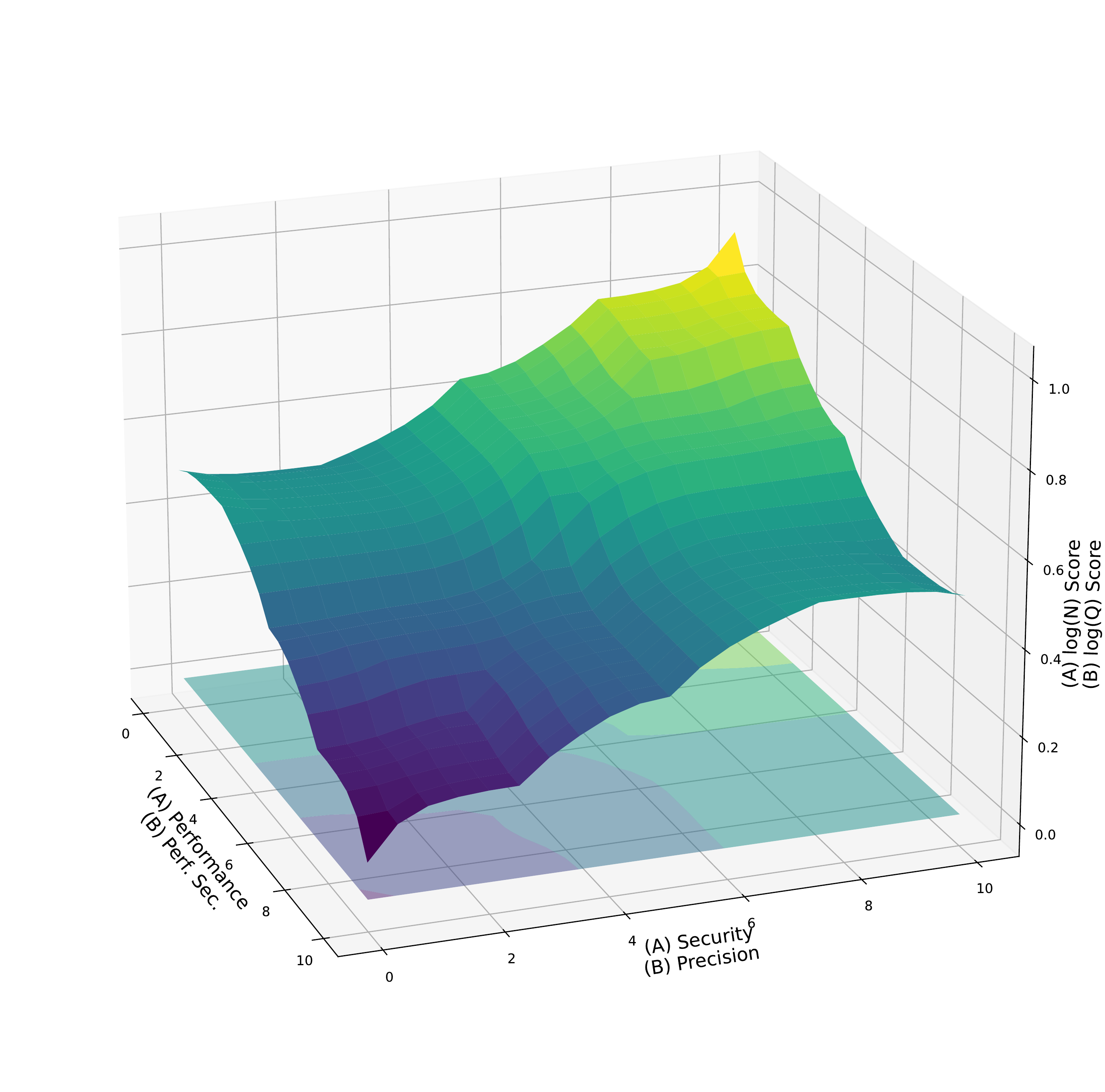}%
    \caption{Fuzzy Logic used for estimating coefficients to weight the polynomial degree $N$ and polynomial modulus $Q$ ($k_{logN}$ and $k_{logN}$). With (A), we represent the antecedents and consequent of the FIP to obtain $k_{logN}$. With (B), we represent the antecedents and consequent of $k_{logQ}$.}
    \label{fig:logNQ_coeff}%
\end{figure}

We use five membership functions for the antecedent and consequent values \textit{very low}, \textit{low}, \textit{medium}, \textit{high} and \textit{very high}. Figure~\ref{fig:logNQ_coeff} represents the FIP that outputs the two coefficients $k_{logN}$ and $k_{logQ}$. The shape achieved by this fuzzy logic allows us to prioritize medium or close to extreme values, where we see plateaus. In general, anything that does not lie within the medium category shifts to them (i.e., we see the local maxima on those points).

As a result of the two Fuzzy Logic modules, the proposed system outputs four coefficients: $k_{real}, k_{dec}, k_{logN}$ and $k_{logQ}$. In the following sections, the variables are used to weight and define various ranges of it.

\subsection{Linear Programming Tasks}
LP solvers intend to provide an optimal solution for a problem defined in an LP model. In our case, the problem is to generate a valid parametrization for $N$ and $Q$. \footnote{As we described in Section~\ref{subsec:lwe_parametrization}, normally, the value of $\sigma$ is fixed to 3.2 due to the reduced increase in noise of that distribution. As such, we do not include this parameter in the selection.}. The LP model receives inputs from the FL modules, some of which are used to define Global Parameters, which are presented first. Then, we describe the different characteristics of the LP model, i.e., the decision variables, the objective function, and the constraints.

\subsubsection{Global Parameters}
As a result of the two Fuzzy Logic modules, the system outputs four coefficients $k_{real}, k_{dec}, k_{logN}$ and $k_{logQ}$. 
The use of these coefficients require some transformations for their integration in the LP model.  
Concretely, we define global parameters for the LP model which define the range of allowed values for $logQ$ and the precision $p$ (i.e., the real scale). 
The system performs it according to the maximum ($max(\mathcal{O}_\mathcal{D}^N)$) and minimum ($min(\mathcal{O}_\mathcal{D}^N$) possible multiplication depths. Concretely, the initial interval is described as:
$$[logQ_{min} = k_{dec}^{min(\mathcal{O}_\mathcal{D}^N)} , logQ_{max} = k_{dec}^{max(\mathcal{O}_\mathcal{D}^N)}]$$ 
$$[p_{min} = k_{real}^{min(\mathcal{O}_\mathcal{D}^N)} , p_{max} = k_{real}^{max(\mathcal{O}_\mathcal{D}^N)}]$$ 
Limiting the range to $[logQ_{min}, logQ_{max}$] turns into a very static model, where the LP solver has little room to optimize the parameters. As such, and based on the precision-performance-security tradeoff, we extend the interval by a factor of $k_{logQ}$ such that: 
$$logQ_{min}^{ext} = logQ_{min} + logQ_{min} \cdot k_{logQ}$$ 
$$logQ_{max}^{ext} = logQ_{max} + logQ_{max} \cdot k_{logQ}$$

\subsubsection{Decision Variables}
The LP task uses five sets of variables. Some of these are auxiliary because they do not have a predefined term in the objective function. The solvers set the values for these auxiliary variables by relating them with other variables in the constraints. We model the problem with auxiliary decision variables since LP tasks cannot express quadratic relations between variables. With the auxiliary variables, we can model the connection between a variable in the objective function and the auxiliary variable in the constraints. 

\begin{description}
    \item[$b^{choice}_{N, \lambda, \mathcal{T}}$] are a set of boolean decision variables that indicate the election of a given $N$. For a given $N$, there are two additional parameters which determine the maximum budget for the $Q$ bit length ($\mathcal{B}_{N, \lambda, \mathcal{T}}$). These parameters are, i) the type of security ($\mathcal{T} \in \{classical, quantum\}$) and, ii) the security parameter ($\lambda \in \{128, 192, 256\}$). Thus, for the model to choose any option, we generate a boolean variable for each combination of a given $N$ under a type of security $\mathcal{T}$ and security parameter $\lambda$, which indicates the election, or not, of that $b^{choice}_{N, \lambda, \mathcal{T}}$ set. 
    More formally, the set of variables is defined as $b^{choice}_{N, \lambda, \mathcal{T}} \in \mathbb{Z}_{2} \forall N, \lambda ,\mathcal{T}$.

    \item[$z_{logq_i}$] are integer auxiliary decision variables that determine the bit length of each element in the polynomial moduli chain. Since there are different $\mathcal{O}_\mathcal{D}^N$ for each $N$, we create  as many as the maximum $max(\mathcal{O}_\mathcal{D}^N)$ of decision variables. Then, our LP model and other variables ensure the use of the minimum required number of $logq_i$ and that this number coincides with the particular multiplication depth $\mathcal{O}_\mathcal{D}^N$ for that $N$. 
    
    \item[$b_{logq_i}^{set}$] are a set of boolean variables used to define whether the variable $z_{logq_i}$ has a value or not. This allows us to properly account for $z_{logq_i}$ during the calculation of the specific budget $\mathcal{B}_{N, \lambda, \mathcal{T}}$ for a polynomial degree selection $b^{choice}_{N, \lambda, \mathcal{T}}$.
    
    \item[$b_{logq_i}^{thr}$] are a set of boolean variables used to define whether the corresponding $z_{logq_i}$ has surpassed a certain threshold. We combine these variables with the global parameters obtained from the Looseness FL Module to enable flexibility of the parameter choice. Furthermore, it also avoids setting values far from the initial user choice.
    
    \item[$z_{p}$] is an integer auxiliary decision variable used to define the real scale precision. This value is left for the LP task to be chosen within the range $[p_{min}, p_{max}]$. The variable combines with $z_{logq_i}$ so that at least a level of precision is guaranteed. Contrary to $z_{logq_i}$, with the real precision, we want to maintain precision from there that the intervals are narrower. 
\end{description}

\subsubsection{Objective Function}
The objective function defines the metrics that need to be minimized or maximized. In this work, we define a cost for every decision variable, and the objective is to minimize it. As discussed in Section~\ref{sec:background-he}, there is no objective way to determine whether an HE parametrization is more suitable before execution. We rather compare them in terms of \textit{performance}, \textit{precision} and \textit{security}. Due to that, we perform a multistep quantification of each decision variable. 
First, we establish the metrics used to compare the different parameter selections. Second, we normalize the metrics in the $[0, 1]$ range and combine them by aggregate multiplying all in a final metric. The resulting metric scores the decision variables in the objective function.  
Overall, $N$ and $Q$ share an equivalent weight in the minimization of the objective function.

\noindent \textbf{Polynomial Degree $N$ Metrics}. To properly account for the impact of the polynomial degree, we define seven different metrics. Table~\ref{table:poly_n_metrics} shows the definition and description of these metrics. To unify the impact of a metric $m$, we normalize every metric in the $[0, 1]$ range with $C_{m'} = \dfrac{C_{m} - min(C_{m})}{max(C_{m}) - min(C_{m})} $. Finally, all metrics are combined with the multiplication and renormalized as in $C_{\Pi} = C_{\lambda'} \cdot C_{N'} \cdot  C_{\mathcal{O}_{add}'} \cdot C_{\mathcal{O}_{mul}'} \cdot C_{\mathcal{O}_{rot}'} \cdot C_{\mathcal{T}'} \cdot C_{|v|_{max}'}$. With this processing we achieve a uniformity on the impact of every metric over $b^{choice}_{N, \lambda, \mathcal{T}}$. In that way, the objective function takes a uniform and equal representation of the different possible choices, taking a value of $0$ for the best and $1$ for the worst. 

\noindent \textbf{Polynomial Modulus $Q$ Metrics}. To loosen the linear programming task when choosing the values for $Q$, we introduce only the boolean variables $b_{logq_i}^{set}$ and $b_{logq_i}^{thr}$ in the objective function. The $b_{logq_i}^{set}$ variables penalize the objective function with each use, promoting the use of less $z_{logq_i}$ whenever possible. We ensure that the amount of $b_{logq_i}^{set}$ variables needed are chosen with the constraints. Also, with the constraints we link $z_{logq_i}$ decision variables with $b_{logq_i}^{set}$ and $b_{logq_i}^{thr}$. In that way, we force $z_{logq_i}$ to take a value within $[logQ_{min}, logQ_{max}^{ext}]$. Also, depending on the value of $k_{logQ}$ we use the decision variable $b_{logq_i}^{thr}$ to reward (i.e., $k_{logQ} \leq 0.5$) or penalize (i.e., $k_{logQ} > 0.5$) the variable $z_{logq_i}$ if it is set above the range $logQ_{min}^{ext}$. In such a way, we promote choosing higher values of $z_{logq_i}$ when precision is the goal and lower values of $z_{logq_i}$ when performance or security are the goal. In all cases, the weight of $b_{logq_i}^{set}$ and $b_{logq_i}^{thr}$ in the objective function is $\dfrac{1}{max(\mathcal{O}_{\mathcal{D}}^N)}$.

{\renewcommand{\arraystretch}{2}
\begin{table}[ht]
\centering
\setlength\tabcolsep{2pt}
\begin{threeparttable}
\fontsize{9}{10.5}\selectfont
\begin{tabular}{c|cc|c}
\hline
\hline
\multicolumn{4}{c}{Polynomial Degree $N$ Metrics} \\
\hline
Metric & \multicolumn{2}{c|}{Definition} & Formula \\
\hline
\multirow{1}{*}{$C_\lambda$} & \multicolumn{2}{p{.5\columnwidth}|}{Establishes the security guarantees of the current $N$.} & $1 - \dfrac{\lambda_i - \lambda_{min}}{\lambda_{max} - \lambda_{min}}$ \\ \hline

$C_N$ & \multicolumn{2}{p{.5\columnwidth}|}{Establishes the tradeoff between performance and security obtained by choosing a particular $N$ considering $k_{logN}$.} & $a = \dfrac{N - N_{min}}{N_{max} - N_{min}}$\\ 
& & & $(1 - k_{logN}) \cdot a + (1 - a) \cdot k_{logN}$\\ \hline

$C_{\mathcal{O}_{add}}$ & \multicolumn{2}{p{.5\columnwidth}|}{Defines the cost of performing additions given the specific circuit and multiplication depth $\mathcal{O}_\mathcal{D}^N$} & $\dfrac{\mathcal{O}_{add} - \mathcal{O}_{add}^{min}}{\mathcal{O}_{add}^{max} - \mathcal{O}_{add}^{min}}$ \\ \hline

$C_{\mathcal{O}_{mul}}$ & \multicolumn{2}{p{.5\columnwidth}|}{Defines the cost of performing multiplications given the specific circuit and multiplication depth $\mathcal{O}_\mathcal{D}^N$} & $\dfrac{\mathcal{O}_{mul} - \mathcal{O}_{mul}^{min}}{\mathcal{O}_{mul}^{max} - \mathcal{O}_{mul}^{min}}$ \\ \hline

$C_{\mathcal{O}_{rot}}$ & \multicolumn{2}{p{.5\columnwidth}|}{Defines the cost of performing rotations given the specific circuit and multiplication depth $\mathcal{O}_\mathcal{D}^N$} & $\dfrac{\mathcal{O}_{rot} - \mathcal{O}_{rot}^{min}}{\mathcal{O}_{rot}^{max} - \mathcal{O}_{rot}^{min}}$ \\ \hline

$C_\mathcal{T}$ & \multicolumn{2}{p{.5\columnwidth}|}{Equalizes the cost of choosing a quantum-safe and a classical parametrization.} & $\dfrac{\mathcal{B}_{N, \lambda, \mathcal{T}}}{\mathcal{B}_{N, \lambda, \mathcal{T}=quantum}}$ \\\hline

$C_{|v|_{max}}$ & \multicolumn{2}{p{.5\columnwidth}|}{Measures the amount of ciphertext vectors needed to represent the given maximum lenght vector $|v|_{max}$ under the current degree $N$.} & $\ceil{\dfrac{N}{2 \cdot |v|_{max}}}$ \\\hline

\end{tabular}
\caption{Metrics used in Linear Programming to evaluate the suitability of a polynomial degree $b^{choice}_{N, \lambda, \mathcal{T}}$ according to the different areas of influence it has. For each metric, a value from 0 to 1 is extracted, with 1 being the worst and 0 being the best. The different values are always weighted from 0 to 1 before being multiplied in a final metric $C_\Pi$. 
}\label{table:poly_n_metrics}
\end{threeparttable}
\end{table}
}

\subsubsection{Constraints}
Finally, after defining the decision variables and objective function, we introduce the constraints that guarantee to fulfill the requirements established in the HE Standard~\cite{he_standard}
Again, we note the constraints ensure that the resulting parametrization of the linear programming task is always within the needed security standards. Although we do not note them as explicit constraints, all decision variables must take a positive value in linear programming. Also, we remind that $z_{logq_i}$ and $z_{p}$ are integer variables and $b^{choice}_{N, \lambda, \mathcal{T}}$, $b_{logq_i}^{set}$ and $b_{logq_i}^{thr}$ are boolean variables.

\begin{description}
    \item[Variable Ranges.] We set constraints that define the ranges for each variable. In this case, we set the variable $z_{logq_i}$ to lie in the range $[0, 60]$. Also, the variable $z_p$ is constrained in the interval $[p_{min}, p_{max}]$ so that the value of $z_p$ is congruent with the precision established in the FL modules.
    \item[Choose one $N_{choice}^{N, \lambda, \mathcal{T}}$]. This constraint guarantees that a single choice is made for $N$, $\lambda$, and $\mathcal{T}$. In that way, the LP task is forced to set to one a single $b^{choice}_{N, \lambda, \mathcal{T}}$ and the rest of $b^{choice}_{N, \lambda, \mathcal{T}}$ to 0 such as:
    $$\sum^{N}_i \sum^{\lambda}_j \sum^{\mathcal{T}}_k b^{choice}_{i,j,k} = 1$$
    
    \item[Adjust to the budget $\mathcal{B}_{N, \lambda, \mathcal{T}}$.] The constraint guarantees that, for each $b^{choice}_{N, \lambda, \mathcal{T}}$, we do not surpass the associated budget with the different variables $z_{logq_i}$.
    $$\sum^{N}_i \sum^{\lambda}_j \sum^{\mathcal{T}}_k b^{choice}_{i,j,k} \cdot \mathcal{B}_{i, j, k} = \sum_i^{\mathcal{O}_\mathcal{D}} z_{logq_i}$$

    \item[Guarantee enough rescaling for the circuit.] This constraint ensures the availability of enough moduli for each multiplication depth. It assures that all rescalings can take place after multiplication. More formally, we want to ensure that for an $N$, there are at least $\mathcal{O}_\mathcal{D}^N$ moduli $b_{logq_i}^{set}$ for relinearization.
    $$\sum^{N}_i \sum^{\lambda}_j \sum^{\mathcal{T}}_k b_{choice}^{i,j,k} \cdot \mathcal{O}_\mathcal{D}^i = \sum_i^{\mathcal{O}_\mathcal{D}} b_{logq_i}^{set}$$
    \item[Guarantee precision.] This constraint ensures there are at least $z_{p}$ bits of precision to represent the real scale. We remind the reader that the previous constraint set the interval for minimum and maximum values for $z_{p}$. It performs the constraining by setting the bit distance $z_{p}$ between the special prime $z_{logq_0}$ and the rest of the moduli ($z_{logq_i}~|~ 0 < i < \mathcal{O}_\mathcal{D}^N$).
    $$z_{logq_i} + p_{choice} \leq z_{logq_0} \forall i \in {0 < i < max(\mathcal{O}_\mathcal{D}^N)}$$
    \item[Precise encryption.] This constraint guarantees that the encryption provides enough bits of precision. We set $z_{logq_0}$ to be the same as $z_{logq_{\mathcal{O}_\mathcal{D}^N}}$.
    $$z_{logq_0} = z_{logq_{\mathcal{O}_\mathcal{N}}}$$

    \item[Pair $z_{logq_i}$ with $b_{logq_i}^{set}$] guarantees that if $b_{logq_i}^{set}$ is 1, then the related value of $logq_i$ has to be within the expected range defined by $[logQ_{min}, logQ_{max}^{ext}]$. Also, if the value of $b_{logq_i}^{set}$ is 0, then the value of $z_{logq_i}$ is necessarily 0, thus not influencing all the previous constraint computations.
    $$z_{logq_i} \geq b_{logq_i}^{set} \cdot logQ_{min} \forall i \in  0 < i < max(\mathcal{O}_\mathcal{D}^N)$$
    $$z_{logq_i} \leq b_{logq_i}^{set} \cdot logQ_{max}^{ext} \forall i \in 0 < i < max(\mathcal{O}_\mathcal{D}^N)$$

    \item[Promote or discourage higher precision values.] \sloppy The LP task has certain flexibility to choose $z_{logq_i}$ based on user choices. If the objective function is positive for $b_{logq_i}^{thr}$, then this constraint will try to preserve values in the range $[logQ_{min}, logQ_{min}^{ext}]$. That means it will look for performance and security rather than precision. Otherwise, if the objective function is negative for $b_{logq_i}^{thr}$  then, we look for values in $[logQ_{min}^{ext}, logQ_{max}^{ext}]$ though it is not enforced.. The constraint is formulated in any case as:
    $$z_{logq_i} \geq b_{logq_i}^{thr} \cdot logQ_{min}^b \forall i \in 0 < i < max(\mathcal{O}_\mathcal{D}^N)$$
\end{description}

\section{Evaluation}
\label{sec:evaluation}
In this section, we evaluate the correctness of the parameter selection of the proposed system. First, we explain the experimental design conducted and then we present the results.

\subsection{Experimental design}
The proposed expert system selects parameters based on i) the user choices for the priority of the precision, performance and security, and ii) the processed circuit (i.e., multiplication depth and maximum vector length). 
The experimental design aims to provide insights into the correctness of the system. For that, we create seven tests that apply different computations and imply different aspects of HE operations. Concretely, the tests evaluate encryption and decryption, additions, multiplications, rotations, rescaling, and combinations of the previous operations. Table~\ref{tab:tests} details the different tests executed. 

\begin{table}
\centering
\setlength\tabcolsep{2pt}
\begin{threeparttable}
\fontsize{9}{10.5}\selectfont
\begin{tabular}{c|ccc}
\hline
\hline
\multicolumn{4}{c}{Benchmark Tests} \\
\hline
Test & \multicolumn{3}{c}{Details} \\
\hline
T1 & \multicolumn{3}{p{.9\columnwidth}}{\textbf{Encrpytion Decryption Test} performs the encryption and decryption of a packed ciphertext.} \\ \hline
T2 & \multicolumn{3}{p{.9\columnwidth}}{\textbf{Addition Test} performs a large number of additions on packed ciphertexts.} \\ \hline
T3 & \multicolumn{3}{p{.9\columnwidth}}{\textbf{Multiplication Test} performs a large number of multiplications on packed ciphertexts with multiplication depth $\mathcal{O}_\mathcal{D}^N = 1$.} \\ \hline
T4 & \multicolumn{3}{p{.9\columnwidth}}{\textbf{Rotation Test} performs a large number of rotations on packed ciphertexts.} \\ \hline
T5 & \multicolumn{3}{p{.9\columnwidth}}{\textbf{Large Multiplication Depth ($\mathcal{O}_\mathcal{D}^N = 5$) Test} performs a large number of multiplications at different multiplication depths up to the maximum.} \\ \hline
T6 & \multicolumn{3}{p{.9\columnwidth}}{\textbf{Larger Multiplication Depth ($\mathcal{O}_\mathcal{D}^N = 10$) Test} performs a large number of multiplications at different multiplication depths up to the maximum.} \\ \hline
T7 & \multicolumn{3}{p{.9\columnwidth}}{\textbf{Matrix Multiplication Test} performs a large number of matrix multiplications between a packed vector and a cleartext matrix which is vectorized according to the diagonal matrix multiplication~\cite{Cabrero2022Towards}.} \\ \hline
\hline

\end{tabular}
\caption{Benchmark tests performed to evaluate the different parameter selections of the expert system. The different algorithms are executed for each user choice value.  
}\label{tab:tests}
\end{threeparttable}
\end{table}

To evaluate \textbf{precision}, we measure the noise introduced by HE operations. First, we use the expert system to extract parameters for different values for priorities of the precision. Second, we execute the benchmark tests in plaintext and using HE. Third, we compare the expected plaintext result with the HE result to measure the error incurred (if any). In order to make the evaluation more accurate, we profit from packing, which allows us to introduce a vector of numbers per ciphertext operating over all of them in SIMD, making the evaluation more uniform. For more details on algorithm creation with packing, we refer the reader to ~\cite{Cabrero2022Towards}.

To evaluate \textbf{performance}, we use the expert system to generate parameters for various values of priorities for the performance. Then, we execute the program measuring the encryption, processing, and decryption time. Note that, for consistency, we force the rescaling or level dropdown in the evaluation. The rationale is that if the set of chosen parameters introduces less noise, then rescaling may be unnecessary, introducing an inequality for the remainder of the test in the number of operations performed in each test and their cost.

To evaluate \textbf{security}, we assess the parameter sets for different values of priorities for the security parameter. As we described in Section~\ref{sec:background-he}, as $logN$ increases, it provides more security, but as $logq_i$ increases, it reduces the security. Accordingly, we theoretically evaluate by analyzing the chosen parameters $logN$ and $logq_i$ and taking into account the security guidelines of the HE Standard~\cite{he_standard}. 

We conduct two different experiments with two types of parametrizations. First, we analyze each variable (performance, security, and accuracy) individually, and show different values for one metric that affects the overall system. For that, we fix the priority value for two metrics at their lowest (i.e., 0) and set the third to values ranging from 0 to 10. This allows to observe the output for the different values for each variable when the others do not affect the result. 
The second experiment attempts to evaluate how the system behaves with conflicting values of the variables, using intermediate priorities. Concretely, we take 4 and 9 as possible priorities for each variable, and run the benchmark tests on all the possible combinations for these two values.  

During the experimentation, each individual test is executed ten times, and the average of the executions is reported. We run the tests on a machine Intel Xeon E5-2683 v3 @ 2.00GHz with 56 cores and 64 GB of RAM running CentOS8. The tests are run on Golang 1.16 and the HE framework Lattigo v2~\cite{lattigo}.

\subsection{Results}

\begin{figure*}[!htbp]
    \centering
    \includegraphics[width=\textwidth]{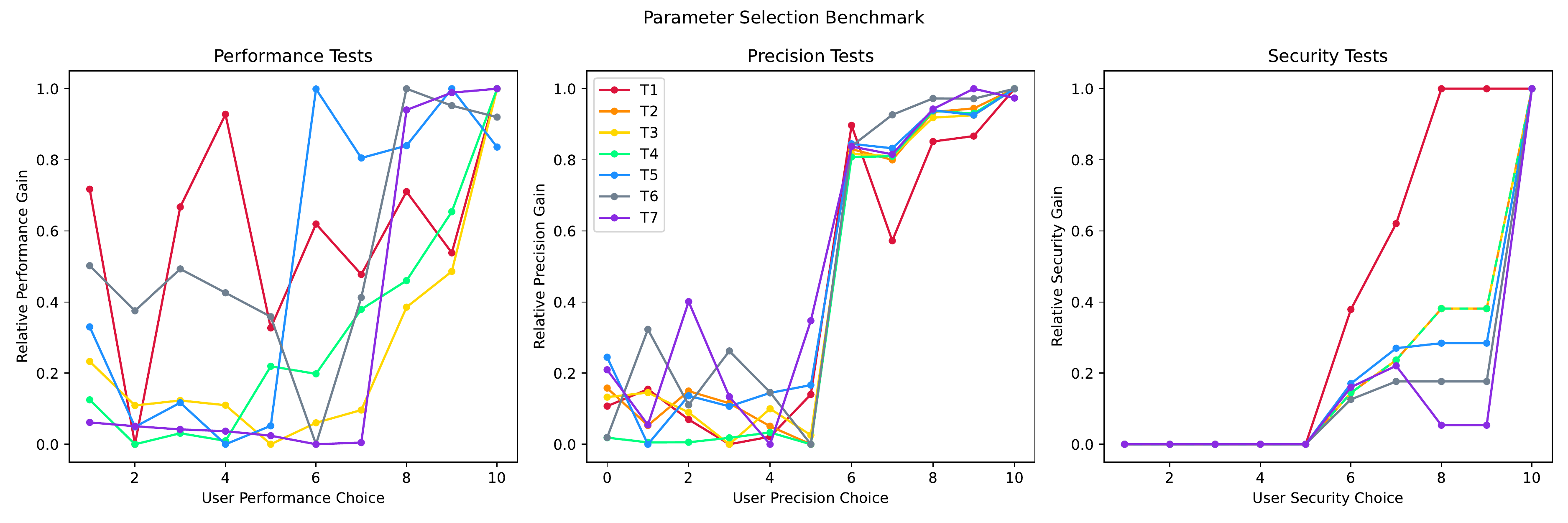}
    \caption{Combined representation of the results relatively aggregated. The figures show the relative improvements in performance, precision and security from generating parameter selections with the expert system. In all these tests, we scale the figures so that, a higher metric means a better result, in all terms, performance, security and performance.}
    \label{fig:plots}%
\end{figure*}

\begin{table*}
\centering
\setlength\tabcolsep{2pt}
\addtolength{\leftskip} {-3cm}
\addtolength{\rightskip}{-3cm}
\begin{threeparttable}
\fontsize{9}{10.5}\selectfont
\begin{tabular}{c|c|c||ccccc||ccccc||ccccc}
\hline
\hline
\multicolumn{18}{c}{Benchmark Tests} \\
\hline
\multirow{2}{*}{Prec.} & \multirow{2}{*}{Perf.} & \multirow{2}{*}{Sec.} & \multicolumn{5}{c||}{T1} & \multicolumn{5}{c||}{T2} & \multicolumn{5}{c}{T3} \\ 
 &  &  & $\Delta$ Res.& Time(s) & $logN$ & $logQ$ & Scale &  $\Delta$ Res.& Time(s) & $logN$ & $logQ$ & Scale & $\Delta$ Res.& Time(s) & $logN$ & $logQ$ & Scale  \\ \hline \hline
4 & 4 & 4 & 1.53e-01 & 8.89e-03 & 12 & 93 & 21 & 2.29e-01 & 4.13e+00 & 12 & 93 & 21 & 4.31e-01 & 1.81e+01 & 12 & 93 & 21\\ \hline
4 & 4 & 9 & 6.45e-01 & 8.52e-03 & 12 & 82 & 19 & 9.11e-01 & 4.14e+00 & 12 & 82 & 19 & 2.16e+00 & 1.81e+01 & 12 & 82 & 19\\ \hline
4 & 9 & 4 & 6.02e-01 & 8.62e-03 & 12 & 82 & 19 & 8.57e-01 & 4.12e+00 & 12 & 82 & 19 & 1.78e+00 & 1.80e+01 & 12 & 82 & 19\\ \hline
4 & 9 & 9 & 5.68e-01 & 8.81e-03 & 12 & 82 & 19 & 9.41e-01 & 4.10e+00 & 12 & 82 & 19 & 1.63e+00 & 1.78e+01 & 12 & 82 & 19\\ \hline
9 & 4 & 4 & 3.40e-02 & 8.78e-03 & 12 & 109 & 23 & 5.88e-02 & 4.12e+00 & 12 & 109 & 23 & 1.17e-01 & 1.80e+01 & 12 & 109 & 23\\ \hline
9 & 4 & 9 & 1.55e-01 & 9.01e-03 & 12 & 93 & 21 & 2.32e-01 & 4.12e+00 & 12 & 93 & 21 & 5.33e-01 & 1.80e+01 & 12 & 93 & 21\\ \hline
9 & 9 & 4 & 1.38e-01 & 8.85e-03 & 12 & 93 & 21 & 2.06e-01 & 4.12e+00 & 12 & 93 & 21 & 3.95e-01 & 1.79e+01 & 12 & 93 & 21\\ \hline
9 & 9 & 9 & 1.37e-01 & 8.88e-03 & 12 & 93 & 21 & 2.25e-01 & 4.12e+00 & 12 & 93 & 21 & 4.14e-01 & 1.80e+01 & 12 & 93 & 21\\ \hline \hline
\multirow{2}{*}{Prec.} & \multirow{2}{*}{Perf.} & \multirow{2}{*}{Sec.} & \multicolumn{5}{c||}{T4} & \multicolumn{5}{c||}{T5} & \multicolumn{5}{c}{T6} \\ 

 &  &  & $\Delta$ Res.& Time(s) & $logN$ & $logQ$ & Scale &  $\Delta$ Res.& Time(s) & $logN$ & $logQ$ & Scale & $\Delta$ Res.& Time(s) & $logN$ & $logQ$ & Scale  \\ \hline \hline
4 & 4 & 4 & 1.47e+00 & 2.53e+01 & 12 & 93 & 21 & 1.15e+02 & 1.01e+01 & 13 & 198 & 21 & 2.28e+03 & 1.42e+02 & 14 & 300 & 21\\ \hline
4 & 4 & 9 & 3.01e+00 & 2.54e+01 & 12 & 82 & 19 & 4.77e+02 & 1.00e+01 & 13 & 174 & 19 & 2.10e+06 & 1.36e+02 & 14 & 280 & 19\\ \hline
4 & 9 & 4 & 2.96e+00 & 2.53e+01 & 12 & 82 & 19 & 5.43e+02 & 1.02e+01 & 13 & 174 & 19 & 2.10e+06 & 1.38e+02 & 14 & 280 & 19\\ \hline
4 & 9 & 9 & 2.91e+00 & 2.51e+01 & 12 & 82 & 19 & 5.08e+02 & 1.01e+01 & 13 & 174 & 19 & 2.10e+06 & 1.37e+02 & 14 & 280 & 19\\ \hline
9 & 4 & 4 & 1.12e+00 & 2.54e+01 & 12 & 109 & 23 & 3.43e+01 & 2.06e+01 & 14 & 232 & 24 & 3.16e+02 & 1.42e+02 & 14 & 352 & 24\\ \hline
9 & 4 & 9 & 1.47e+00 & 2.51e+01 & 12 & 93 & 21 & 1.18e+02 & 1.00e+01 & 13 & 198 & 21 & 2.96e+03 & 1.41e+02 & 14 & 300 & 21\\ \hline
9 & 9 & 4 & 1.47e+00 & 2.50e+01 & 12 & 93 & 21 & 1.17e+02 & 1.01e+01 & 13 & 198 & 21 & 2.45e+03 & 1.42e+02 & 14 & 300 & 21\\ \hline
9 & 9 & 9 & 1.48e+00 & 2.53e+01 & 12 & 93 & 21 & 1.26e+02 & 9.98e+00 & 13 & 198 & 21 & 2.18e+03 & 1.42e+02 & 14 & 300 & 21\\ \hline \hline
\multirow{2}{*}{Prec.} & \multirow{2}{*}{Perf.} & \multirow{2}{*}{Sec.} & \multicolumn{5}{c||}{T7}  \\ 

 &  &  & $\Delta$ Res.& Time(s) & $logN$ & $logQ$ & Scale  \\ \cline{1-8}\cline{1-8}
4 & 4 & 4 & 7.03e+01 & 1.82e+01 & 13 & 136 & 21\\ \cline{1-8}
4 & 4 & 9 & 1.53e+02 & 1.82e+01 & 13 & 120 & 19\\ \cline{1-8}
4 & 9 & 4 & 1.98e+02 & 1.85e+01 & 13 & 120 & 19\\ \cline{1-8}
4 & 9 & 9 & 2.83e+02 & 1.81e+01 & 13 & 120 & 19\\ \cline{1-8}
9 & 4 & 4 & 2.42e+00 & 1.81e+01 & 13 & 160 & 24\\ \cline{1-8}
9 & 4 & 9 & 1.12e+02 & 1.81e+01 & 13 & 136 & 21\\ \cline{1-8}
9 & 9 & 4 & 4.29e+01 & 1.82e+01 & 13 & 136 & 21\\ \cline{1-8}
9 & 9 & 9 & 5.68e+01 & 1.81e+01 & 13 & 136 & 21\\ \cline{1-8}
\cline{1-8}

\end{tabular}
\caption{Numeric benchmarks for each test executed on values of precision, performance and security for all combinations of 4 and 9. The columns for each test show $\Delta$ Res. as the variation of the result with the HE noise, with the expected result (i.e., computed in cleartext); the total time in seconds (s) needed to execute the test and the security parameters used $logN$, $logQ$ and scale.  
}\label{tab:evaluation}
\end{threeparttable}
\end{table*}

We divide this section into the analysis of the two experiments used, one to evaluate each feature individually, using all possible priority values, and the other to test the different combinations of conflicting priority.

The results for the individual tests are shown in Figure~\ref{fig:plots}. We note that, due to the nature of the tested circuits, the parametrizations may not change significantly, and thus the implication of user choices in performance are more irregular. Nevertheless, we observe a clear improvement in the performance gain for higher values of the user choice. For example, in Tests 5, 6, and 7, which are the most complex from our benchmark set, the performance gain is negligible for low values (below 6) and then considerably increases for higher values. 
In the case of precision, there is a clear improvement for values higher than 4 or 5 in all tests, which shows that the expert system successfully materializes the user choices. 
A similar pattern can be observed with security, where in most of the tests, the improvement affects values higher than 5. However, in more complex circuits (i.e., Tests 5, 6, and 7), the increased security is smaller, due to the requirements of the circuit. There is a particular behavior in Test 7 for values 8 and 9, where the security decreases regarding lower priorities. This is due to the performance-security tradeoff, where the LP tasks chose a lower value of $logQ$, and, in doing so, it is able to reduce the budget $\mathcal{B}_{N, \lambda, \mathcal{T}}$ thus fitting the parameters within a smaller $logN$ (and in turn, improving performance). Although the security metric slightly decreases, it remains high according to the HE standard, and we do not consider it as a worse parametrization in practice.

Table~\ref{tab:evaluation} shows the results for the experiment where we have run the test suite using different combinations of priorities for the three features. This experiment evaluates the behavior of the expert system when conflicting user choices appear. We observe that, as a general pattern, when there is no conflict (i.e., all features have the same value, either 4 or 9), the results show that the three features equally improve, obtaining equivalent output. It is mainly due to the Fuzzy Logic membership functions, logic rules, and defuzzification that average and smooths the changes avoiding any decision that has a larger impact over the others. As long as the choices are not diverse, the overall optimization of parameters is shared for the three features. Indeed, we observe that when precision is the most important feature (with a value of 9, versus 4 for the other two variables), the $logQ$ and $scale$ increase, and the error in the result is considerably minimized in all tests. As for security, we observe that for higher values of security choice, the polynomial degree $logN$ is kept high, and the polynomial modulus $logQ$ is minimized. The best performance value is shared among those parametrizations having the lowest $logN$ and $logQ$.

\subsection{Discussion}

From the results of our first experiment, we observe that the improvement in each of the variables increases with higher values of the priority when these are set independently (i.e., the other two variables are not considered). 
When different values are combined and conflict with each other (experiment 2), the expert system also behaves as expected, and the result often favors the variable with higher priority. We observe, however, that in some cases, the different combination of priorities results in the same set of parameters, and turn, in similar accuracy, performance and security. This is expected since the system is designed to guarantee a minimum level of security as defined in the HE Standard. 
Overall, we have shown that the expert system helps to provide optimal values for the parameters, balancing security, performance, and accuracy. Indeed, the system outputs a set of parameters that are congruent with what an expert would initially seek. These choices are performed from the heuristics and expert knowledge implemented within the system. Furthermore, it requires minimal effort from the user to achieve goals by setting high-level parameters. Otherwise, the user would have to manually understand all the constraints and the considerations of each choice. 

\section{Conclusions}

Homomorphic encryption is a technique that enables privacy-preserving computation. Still, its use remains limited due to the complexity of its use. The particularities of algorithm design and parameter selection are among the factors that make HE complex. In this paper, we covered the different aspects of parameter selection and provided an overview of the challenges of parameter selection for a given circuit. Concretely, the existing circular dependencies between the circuit and the parameters make it a complex iterative process, and the conflicting impact of the parameters on the security, performance, and accuracy of the models. 
To address these challenges, we elaborated an expert system combining Fuzzy Logic and Linear Programming. Fuzzy Logic allows users to introduce priority values for precision, performance, and security, which are then processed and transformed to design the Linear Programming model. This model then outputs the selection of parameters that i) preserve the user choices and ii) are optimal for these choices and the circuit considered. Also, 
the system always provides a secure parametrization of the cryptosystem independently of the user choices. 

Our experimental evaluation shows that the underlying parametrizations provided by the expert system work as expected. Furthermore, we demonstrate how choosing precision, performance, or security brings extensive gains in that area. 
We believe that this system considerably contributes to bridge the gap between classic and privacy-preserving computation paradigms, and will foster the application of HE into existing data science frameworks.
Indeed, in future work, extensions of the expert system to elaborate more on the relations between ciphertext scale and rescaling. Also, integrating the system into a global framework to build optimal HE circuits could help it become a better solution.


\section*{Acknolwedgments}
This work was partially supported by CERN openlab, the CERN Doctoral Student Programme, the Spanish grants ODIO (PID2019-111429RB-C21 and PID2019-111429RB) and the Region of Madrid grant CYNAMON-CM (P2018/TCS-4566), co-financed by European Structural Funds ESF and FEDER.
The opinions, findings, and conclusions or recommendations expressed are those of the authors and do not necessarily reflect those of any of the funders.

\bibliographystyle{unsrt}
\bibliography{bib/bib}

\appendix

\section{Membership Function Values}
In Table~\ref{tab:intervals}, we show the different intervals used for membership functions.
\begin{table}
\centering
\setlength\tabcolsep{2pt}
\begin{threeparttable}
\fontsize{9}{10.5}\selectfont
\begin{tabular}{c|ccccc}
\hline
\hline
\multicolumn{6}{c}{Fuzzy Logic Membership Function Interval} \\
\hline
Var. & Very Low & Low & Medium & High & Very High \\
Precision & $[-\infty, 0, 2.5]$ & $[0, 2.5, 5]$ & $[2.5, 5, 7.5]$ & $[5, 7.5, 10]$ & $[7.5, 10, \infty]$ \\
Performance & $[-\infty, 0, 2.5]$ & $[0, 2.5, 5]$ & $[2.5, 5, 7.5]$ & $[5, 7.5, 10]$ & $[7.5, 10, \infty]$ \\
Security & $[-\infty, 0, 2.5]$ & $[0, 2.5, 5]$ & $[2.5, 5, 7.5]$ & $[5, 7.5, 10]$ & $[7.5, 10, \infty]$ \\
$\mathcal{O}_{\mathcal{D}}$ & $[-\infty, 0, 7]$ & $[0, 7, 14]$ & $[7, 14, 21]$ & $[14, 21, 28]$ & $[21, 28, \infty]$ \\
$k_{real}$ & $[-\infty, 0, 7.5]$ & $[0, 7.5, 15]$ & $[7.5, 15, 22.5]$ & $[15, 22.5, 30]$ & $[22.5, 30, \infty]$ \\ \hline \hline
$k_{dec}$ & $[-\infty, 12, 24]$ & $[12, 24, 36]$ & $[24, 36, 48]$ & $[36, 48, 60]$ & $[48, 60, \infty]$ \\
$k_{logN}$ & $[-\infty, 0, 0.25]$ & $[0, 0.25, 0.5]$ & $[0.25, 0.5, 0.75]$ & $[0.5, 0.75, 1.0]$ & $[0.75, 1.0, \infty]$ \\
$k_{logQ}$ & $[-\infty, 0, 0.25]$ & $[0, 0.25, 0.5]$ & $[0.25, 0.5, 0.75]$ & $[0.5, 0.75, 1.0]$ & $[0.75, 1.0, \infty]$ \\
\hline
\hline

\end{tabular}
\caption{Fuzzy Logic Membership Function Intervals for the Triangular functions. Represented as [beggining, peak, end].  
}\label{tab:intervals}
\end{threeparttable}
\end{table}

\end{document}